\documentclass[aps,prb,amsmath,amssymb,twocolumn,superscriptaddress]{revtex4}
\usepackage{graphicx}
\usepackage{braket}
\usepackage{color}
\usepackage{hyperref}
\usepackage{soul}

\newcommand{\MP}{\varepsilon_\mathrm{M}}	
\newcommand{\DP}{\varepsilon_\mathrm{M}^\mathrm{cc}}

\begin{document}

\title{Fast spin exchange between two distant quantum dots}   

\author{Filip K. Malinowski}
\thanks{These authors contributed equally to this work}
\affiliation{Center for Quantum Devices and Station Q Copenhagen, Niels Bohr Institute, University of Copenhagen, 2100 Copenhagen, Denmark}

\author{Frederico Martins}
\thanks{These authors contributed equally to this work}
\affiliation{Center for Quantum Devices and Station Q Copenhagen, Niels Bohr Institute, University of Copenhagen, 2100 Copenhagen, Denmark}

\author{Thomas B. Smith}
\affiliation{Centre for Engineered Quantum Systems, School of Physics, The University of Sydney, Sydney NSW 2006, Australia}

\author{Stephen D. Bartlett}
\affiliation{Centre for Engineered Quantum Systems, School of Physics, The University of Sydney, Sydney NSW 2006, Australia}

\author{Andrew C. Doherty}
\affiliation{Centre for Engineered Quantum Systems, School of Physics, The University of Sydney, Sydney NSW 2006, Australia}

\author{Peter D. Nissen}
\affiliation{Center for Quantum Devices and Station Q Copenhagen, Niels Bohr Institute, University of Copenhagen, 2100 Copenhagen, Denmark}

\author{Saeed Fallahi}
\affiliation{Department of Physics and Astronomy, Station Q Purdue, and Birck Nanotechnology Center, Purdue University, West Lafayette, Indiana 47907, USA}

\author{Geoffrey C. Gardner}
\affiliation{Department of Physics and Astronomy, Station Q Purdue, and Birck Nanotechnology Center, Purdue University, West Lafayette, Indiana 47907, USA}

\author{Michael J. Manfra}
\affiliation{Department of Physics and Astronomy, Station Q Purdue, and Birck Nanotechnology Center, Purdue University, West Lafayette, Indiana 47907, USA}
\affiliation{School of Electrical and Computer Engineering, and School of Materials Engineering, Purdue University, West Lafayette, Indiana 47907, USA}

\author{Charles M. Marcus}
\affiliation{Center for Quantum Devices and Station Q Copenhagen, Niels Bohr Institute, University of Copenhagen, 2100 Copenhagen, Denmark}

\author{Ferdinand Kuemmeth}
\thanks{To whom correspondence should be addressed; E-mail: kuemmeth@nbi.dk.}
\affiliation{Center for Quantum Devices and Station Q Copenhagen, Niels Bohr Institute, University of Copenhagen, 2100 Copenhagen, Denmark}

\maketitle

\date{\today}

{\bf
The Heisenberg exchange interaction between neighboring quantum dots allows precise voltage control over spin dynamics, due to the ability to precisely control the overlap of orbital wavefunctions by gate electrodes. This allows the study of fundamental electronic phenomena \cite{Cronenwett1998, Potok2007, Craig2004, Malinowski2018} and finds applications in quantum information processing\cite{Petta2005}. 
Although spin-based quantum circuits based on short-range exchange interactions are possible~\cite{Watson2018,Zajac2018}, the development of scalable, longer-range coupling schemes constitutes a critical challenge within the spin-qubit community. 
Approaches based on capacitative coupling~\cite{Nichol2017} 
 and cavity-mediated interactions~\cite{Samkharadze2018,Mi2018,Landig2017} effectively couple spin qubits~\cite{Taylor2007,Dial2013} to the charge degree of freedom~\cite{Taylor2005a,Burkard2006}, making them susceptible to electrically-induced decoherence.
The alternative is to extend the range of the Heisenberg exchange interaction by means of a quantum mediator~\cite{Baart2017,Mehl2014a,Srinivasa2015,Croot2017}. 
Here, we show that a multielectron quantum dot with 50-100 electrons serves as an excellent mediator, preserving speed and coherence of the resulting spin-spin coupling while providing several functionalities that are of practical importance.
These include speed (mediated two-qubit rates up to several gigahertz), distance (of order of a micrometer), voltage control, possibility of sweet spot operation~\cite{Martins2016,Reed2016} (reducing susceptibility to charge noise), and reversal of the interaction sign (useful for dynamical decoupling from noise)~\cite{Martins2017,Malinowski2018,Deng2017}.
}

%:fig1
\begin{figure}
	\centering
	\includegraphics[width=0.48\textwidth]{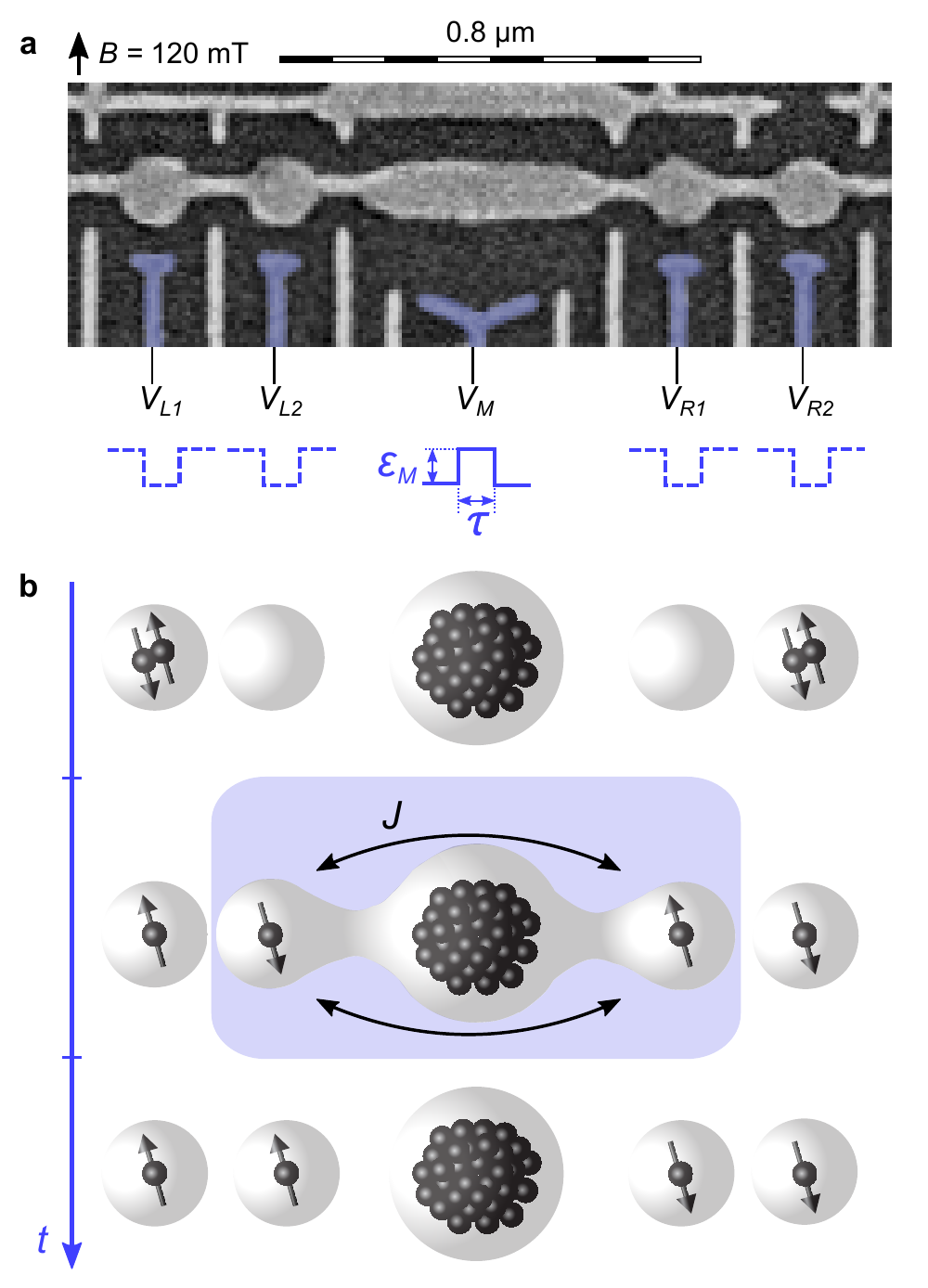}
	\caption{
	{\bf Detection of spin-exchange processes across a multielectron dot.}
	{\bf a} Scanning electron micrograph of the measured device. A multielectron dot is induced below the long segment of the horizontal gate electrode, while two-electron double quantum dots are induced below its circular sections. Nanosecond voltage pulses applied to the blue-colored gates $V_j$ control the position of individual electrons and their mutual interactions.  An external magnetic field (arrow) is applied in-plane of the device.
	{\bf b} Operation steps. First, each double dot is initialized in a singlet state $\ket{S^{L/R}}$, by populating the outer dots with two electrons each. Then, single electrons are moved to the inner dots, thereby turning off their exchange interaction with the outer electrons, which serve as reference spins. Next, the exchange coupling $J$ between the inner electrons is induced, by temporarily raising $V_M$ by an amplitude parameterized by $\MP$ (and lowering other gates to maintain constant overall charge). The exchange interaction causes flip-flops between electronic spins on the inner dots (for mechanisms see Fig.~\ref{fig3}a), which entangles the spin state of the left double dot with that of the right double dot. After an interaction time $\tau$ the resulting correlations in the relative alignment between inner spins and reference spins are detected by spin-to-charge conversion within each double dot, using two nearby sensor quantum dots (not shown). 
	}
	\label{fig1}
\end{figure}

We implement long-range exchange coupling mediated by a multielectron quantum dot in a linear array of five quantum dots, as shown in Fig.~\ref{fig1}a. 
The quintuple dot is defined in a GaAs two-dimensional electron gas by means of electrostatic gate electrodes deposited on top of the heterostructure.
The middle dot is populated by a large even number of electrons, between 50 and 100 as estimated from the lithographic size of the device and the density of the two-dimensional electron gas, and is characterized by a spinless ground state~\cite{Malinowski2018}. 
Two two-electron double dots are tunnel-coupled on opposing sides of the large middle dot and are each initialized and read out using standard techniques for singlet-triplet qubits~\cite{Petta2005}. 

The exchange interaction is induced by a sequence of sub-microsecond voltage pulses applied to the blue-colored gates in Fig.~\ref{fig1}a, realizing the following steps (Fig.~\ref{fig1}b). 
First, the outer dots are each populated by a pair of electrons. This initializes each double dot in the spin singlet state, $\ket{S^{L/R}}=(\ket{\uparrow\downarrow}-\ket{\downarrow\uparrow})/\sqrt{2}$, where arrows indicate the spin state of the two electrons and the superscript $L/R$ indicates the left and right double dot. 
Then, the electron pairs are rapidly separated within each double dot. 
This pulse effectively turns off the exchange interaction within each double dot, allowing the outer dots to each store one reference spin. 
In the third step, $V_M$ is temporarily increased by $\MP$, while negative (compensation) pulses are applied to all other gates (see Methods Section~II). 
This induces an exchange interaction between the inner one-electron dots mediated by the large dot.
After the interaction time $\tau$ the exchange-inducing pulse is switched off. 
Subsequently, spin-to-charge conversion is used to read out the relative spin alignment within each double dot\cite{Petta2005}, independently and with single-shot fidelity.

%:fig2
\begin{figure*}
	\centering
	\includegraphics[width=\textwidth]{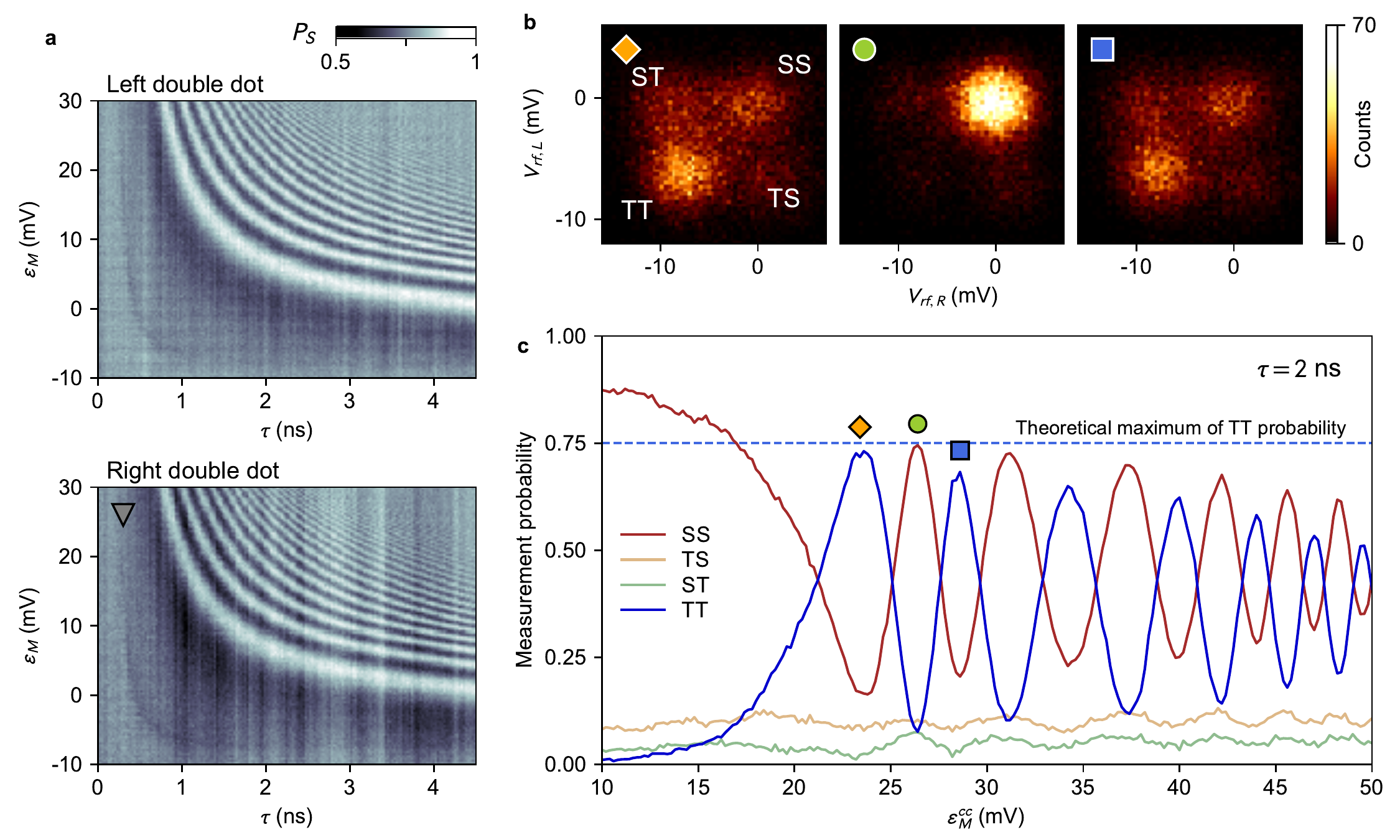}
	\caption{
	{\bf Exchange oscillations across the mediator and non-local correlations.} 
	{\bf a} Fraction of detected singlet outcomes, $P_S$, acquired simultaneously for the left and right double dot, as a function of interaction time $\tau$ and pulse amplitude $\MP$. 
	The choice of detuning between inner dots, $\varepsilon=-2$ mV, corresponds to a symmetric operation point (cf. gray triangle in Fig.~\ref{fig3}b).
    {\bf b} Histograms of demodulated sensor voltages, when repeating a pulse cycle with $\tau=2$~ns many times, for three different choices of $\DP$ as marked in panel c. Counts bunch into four groups, each associated with different combination of a singlet (S) and triplet (T) measurement outcomes of the two double dots. Correlations within these single-shot measurement outcomes reveal the non-local nature of the interaction. 
	{\bf c} Joint probabilities of all four possible joint outcomes, as a function of the exchange-inducing pulse amplitude $\DP$ for fixed interaction time $\tau=2$~ns.
	Here, $\DP$-pulses are defined similar to $\MP$-pulses, but with a different choice for the cross-compensation amplitudes (see Methods Section~II). Dashed line indicates the largest expected probability to detect TT for the maximally entangled state (see text).
	}
	\label{fig2}
\end{figure*}

The result of such a spin-exchange pulse sequence is shown in Fig.~\ref{fig2}a. 
In the two panels, we plot the fraction of singlet outcomes ($P_S$) for each double dot as a function of $\tau$ and $\MP$.
Oscillations in $P_S$ witness exchange-driven flip-flop processes between the two spins located on the inner quantum dots. 
The oscillation frequency increases for larger values of $\MP$. 
Consistent with complementary spin-leakage spectroscopy (see Methods Section~IV), this indicates that positive pulses on gate $V_M$ lower the multielectron dot levels towards resonance with the inner dots, thereby increasing the rate of spin-exchange processes mediated by the multielectron dot.

Next, we demonstrate correlations between measurement outcomes for the left and right double dot. 
For fixed interaction time $\tau = 2$~ns, the demodulated voltage signals for the left and right sensor ($V_{rf,L}$ and $V_{rf,R}$) exhibit correlations that oscillate with the amplitude of the applied pulse~(Fig.~\ref{fig2}b), confirming the non-local mechanism underlying panels~\ref{fig2}a. 
(In Figure~\ref{fig2}b, the exchange-inducing pulses parametrized by $\DP$ are defined similar to $\MP$-pulses in Fig.~\ref{fig2}a, but employ more sophisticated cross-compensation pulses, as described in Methods Section~II). 
From these histograms we extract the joint probabilities of detecting a singlet (S) or triplet (T) for the two double dots as a function of $\DP$ (see Supplementary Video 1 for animated histograms). Figure~\ref{fig2}c clearly shows anticorrelated probabilities for detecting SS and TT, whereas the probabilities of ST and TS are small and nearly constant.
The joint probabilities were extracted by fitting the histograms with four Gaussians and correcting for double-dot relaxation during the measurement pulse (Methods Section~V).

The oscillatory behaviour of joint probabilities result from the precession between the initialized state $\ket{S^L}\ket{S^R}$ and the maximally entangled state $\frac{1}{2}(\ket{S^L}\ket{S^R}-\ket{T_0^L}\ket{T_0^R}+\ket{T_+^L}\ket{T_-^R}+\ket{T_-^L}\ket{T_+^R})$. 
(Here, the two kets denote the state of the left and right double dot, respectively, and the spin triplet states are labeled according to the standard convention, $\ket{T_0}=(\ket{\uparrow\downarrow}+\ket{\downarrow\uparrow})/\sqrt{2}$, $\ket{T_+}=\ket{\uparrow\uparrow}$, $\ket{T_-}=\ket{\downarrow\downarrow}$.) 
The coefficients associated with this entangled state explain the visibility in our measurement basis. For example, the maximum expected probability for TT is 75\%, consistent with the observed maxima in Fig.~\ref{fig2}c. 
The observed visibility for SS and TT is further reduced by residual counts of ST and TS. We attribute this background to unintentional dynamics of the reference spins in the outer dots, arising from decoherence due to their coupling to the nuclear spin bath associated with GaAs~\cite{Malinowski2017a}, and from the finite rise time of the voltage pulses. 

%:fig3
\begin{figure*}
	\centering
	\includegraphics[width=\textwidth]{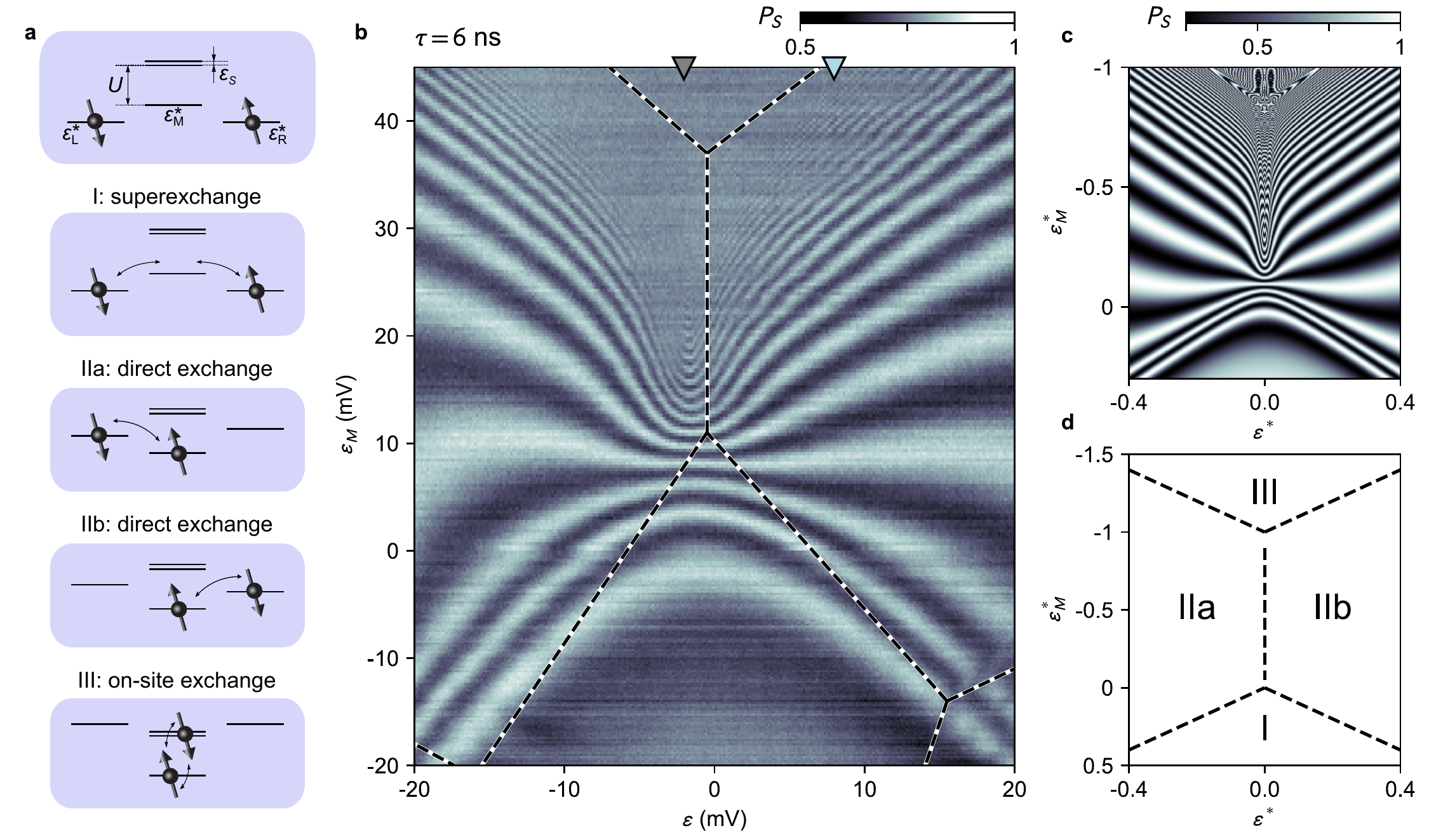}
	\caption{
	{\bf Physical regimes of exchange interaction.}
	{\bf a} In the Hubbard model different spin-exchange processes dominate depending on the relative alignment of various single-particle levels (cf. Methods Section~VII). Specifically, $\varepsilon_M^*$ is the single-particle energy of the lowest unoccupied orbital in the multielectron dot relative to the left and right orbital, $\varepsilon^* = (\varepsilon_L^*-\varepsilon_R^*)/2$ determines the relative detuning between the left and right orbital, and $U$ and $\varepsilon_S$ indicate the charging energy and the level spacing of the multielectron dot. Depending on which processes are energetically allowed or suppressed, we classify different regimes as illustrated.
	{\bf b} Measured $P_S(\varepsilon,\MP)$ for the right double dot for fixed interaction time $\tau=6$~ns. 
	Colored triangles indicate the detuning points used for Fig.~\ref{fig2}a and Fig.~\ref{fig4}b. 
	Dashed lines indicate the location of independently measured charge transitions (Methods Section~VI). 
	{\bf c} Simulated $P_S(\varepsilon^*,\varepsilon_M^*)$  in the Hubbard model.
	{\bf d} Location of charge transitions (dashed lines) in the Hubbard model for the parameters used in c. The corresponding charge configurations of the four regimes of exchange interaction are schematically indicated by dots in a.
	}
	\label{fig3}
\end{figure*}

In Fig.~\ref{fig3} we identify different regimes of the exchange interaction mediated by the multielectron quantum dot. 
For that purpose we define a new gate voltage parameter, $\varepsilon=(V_{L2}-V_{R1})/\sqrt{2} + C$ (where $C$ is a constant; Methods Section~II), which controls the relative detuning between the two inner dots, and plot the readout probabilities $P_S(\varepsilon,\MP)$.
Since $\tau=6$~ns is fixed for these measurements, each fringe in Fig.~\ref{fig3}b corresponds to points of equal exchange energy $J$, while the density of fringes represents the gradient of $J$ (see Methods Section~VIII for a discussion of finite-rise-time effects).
The observed exchange strength increases rapidly for $\MP > $~20~mV, especially for $\varepsilon \approx 0$, resulting in a high density of fringes that is blurred by a combination of aliasing and decoherence. For finite $| \varepsilon |$ the exchange increases more slowly, resulting in a pattern that is approximately symmetric with respect to $\varepsilon$.

The observed pattern can be understood by monitoring the charge distribution during the interaction step (Methods Section~VI), and overlaying the fringes in Fig.~\ref{fig3}b with the observed charge transitions (dashed lines). Consistent with simulations from the Hubbard model discussed below, shown in Fig.~\ref{fig3}c,d, we identify each region with a different charge configuration, as illustrated by dots in Fig.~\ref{fig3}a.
In region I the inner dots remain singly occupied, and the multielectron dot keeps its initial charge state. This corresponds to a superexchange interaction, where virtual tunneling through the multielectron quantum dot mediates the exchange interaction~\cite{Srinivasa2015}.
In regions IIa and IIb one of the electrons has relocated onto the multielectron dot, forming an effective spin-1/2 many-body state which directly exchange-couples to the other electron spin. The mirror symmetry of IIa and IIb with respect to $\varepsilon^*=0$ reflects the left-right symmetry of the device, with minor deviations in the experimental data arising from a slight inequality in the tunneling barriers between the multielectron dot and inner dots.
In region III the chemical potential of the multielectron dot is sufficiently low such that both electrons relocate onto the multielectron dot. Depending on their relative spin alignment, singlet-like or triplet-like, both electrons occupy either the lowest orbital, or the lowest and second lowest orbital, respectively. 
The energy difference between these spin configurations sets the coupling strength of this (rapid) onsite exchange interaction. It is related to two mesoscopic parameters, namely the single-particle level spacing of the two orbitals, and the spin correlation energy~\cite{Malinowski2018,Kurland2000,Folk2001}.

To verify all four regimes we evaluate a Hubbard model of the two inner quantum dots coupled to the multielectron quantum dot (see level structure in Fig.~\ref{fig3}a and Methods Section~VII). Using realistic parameters, this model qualitatively reproduces all observations, including the fringe pattern and the charge stability diagram (Fig.~\ref{fig3}c,d).

%:fig4
\begin{figure*}
	\centering
	\includegraphics[width=\textwidth]{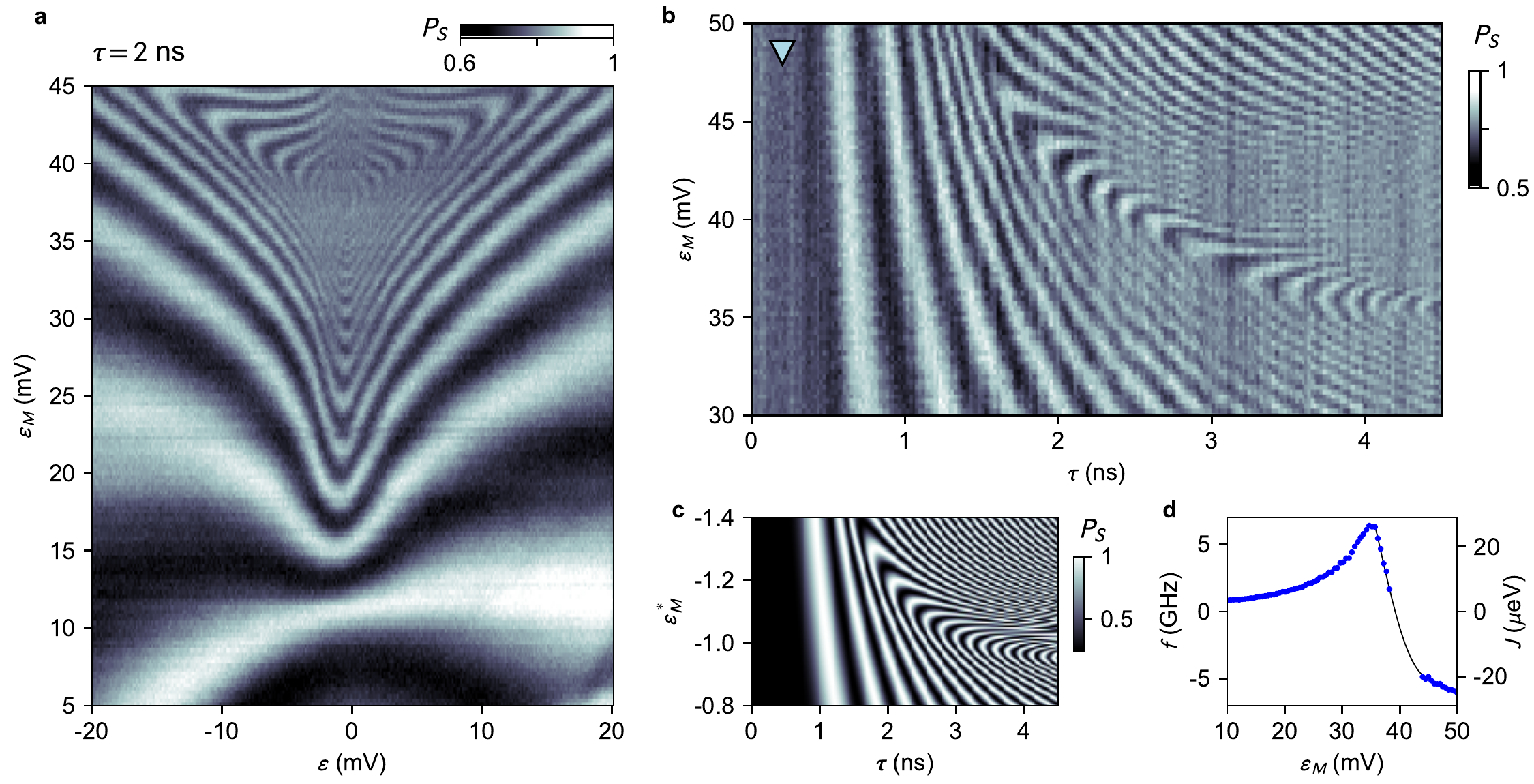}
	\caption{
	{\bf Sweet-spot behavior and competition between direct and onsite processes.}
    	{\bf a} $P_S(\varepsilon, \MP)$ for reduced interaction time $\tau=2$~ns. A fingerprint pattern, related to a sweet spot in the exchange profile $J(\varepsilon,\MP)$, emerges at the crossover from direct to onsite regimes.
	Measured, {\bf b}, and simulated, {\bf c}, time-dependent exchange oscillations for fixed $\varepsilon=8$~mV (marked by blue triangle in Fig.~\ref{fig3}b). Enhanced oscillation visibility along the chevron pattern indicates that operation at the sweet spot prolongs coherence.
	{\bf d} Coupling strength of the quantum mediator, demonstrating high speed, sweet spot, and sign reversal, controlled by small voltage changes in $\MP$. Data points represent the oscillation frequency extracted from rows in panel b, which we identify with the exchange coupling strength $J=hf$. Solid line is a guide to the eye.  
	}
	\label{fig4}
\end{figure*}

Quantitative insight into the fast dynamics of onsite exchange can be obtained by reducing $\tau$ to 2~ns. This circumvents blurring and aliasing effects, revealing a characteristic fringe pattern at the transition between direct and onsite exchange regimes (Fig.~\ref{fig4}a).
This pattern is in fact a fingerprint of the exchange profile $J(\varepsilon,\MP)$~\cite{Reed2016}: Retaking any pixel, say along a cut at $\varepsilon=8$~mV (blue triangle in Fig.~\ref{fig3}b), as a function of $\tau$ results in oscillations with frequency $f=J/h$~\cite{Martins2016}. Extracting $f$ for the cut shown in Fig.~\ref{fig4}b reveals a non-monotonic behavior of the exchange coupling with respect to $\MP$ (Fig.~\ref{fig4}d). The presence of a maximum in frequency followed by a zero crossing is similar to exchange profiles studied in Refs.~\onlinecite{Martins2017,Malinowski2018}, and arises if direct exchange (which depends on orbital-specific tunnel matrix elements) competes with onsite exchange (which depends on electron correlation effects and, for relatively small orbital level spacing, can be negative). 
Accordingly, to qualitatively reproduce the chevron pattern of Fig.~\ref{fig4}b, we must include two unoccupied orbitals of the multielectron quantum dot~\cite{Malinowski2018,Deng2017}, as well as a finite rise time of the applied voltage pulses (Methods Section~VII and VIII).

Furthermore, the visibility of oscillations in panel~\ref{fig4}b depends on $\MP$, which we associate with an enhancement of fidelity in two operating regimes. First, for large values of $\MP$, the onsite exchange energy is set by the (mesoscopic) level spacing of the dot, which to lowest order is insensitive to pulse amplitudes. This regime is akin to the noise-insensitive regimes noted in Refs.~\onlinecite{Dial2013,Medford2013} and exploited by the three-electron double-dot hybrid qubit~\cite{Kim2014,Cao2016}.
Second, high-fidelity oscillations appear along the curved chevron pattern, suggesting that the local extremum in the exchange strength provides insensitivity~\cite{Martins2016,Reed2016} to fluctuations in $\MP$.
For this tuning of the device, the oscillation frequency in both noise-insensitive regimes exceeds 5~GHz, making it challenging to perform small angle rotations using conventional pulse generators. However, by decreasing the tunnel couplings between the multielectron dot and the inner dots the operating speed at the sweet spot can be reduced as needed (down to 1~GHz as demonstrated in Ref.~\onlinecite{Martins2017}).

An interesting next step building upon this demonstration is to employ a multielectron quantum dot of larger dimensions, with multiple single-electron quantum dots around its perimeter. 
This will allow coherent coupling of arbitrary pairs of electrons, and may lead to a programmable hardware architecture in which qubit-qubit connectivities can be reconfigured in situ to best serve the specific computational tasks. 
Increasing the coupler size has additional advantages, such as reducing the onsite exchange energy which would enable performing high-fidelity, small-angle rotations. Another direction is the implementation of this coupling scheme in silicon nanostructures, mitigating decoherence effects arising from the nuclear spin bath. Our demonstration of coherently swapping spin pairs across the multielectron quantum dot suggests that shuttling of individual electrons~\cite{Fujita2017} through the multielectron quantum dot should also be feasible. Combinations of these achievements will open many paths for scaling quantum-dot based qubit circuits.

\section*{Acknowledgments}
This work was supported by the Army Research Office, the Innovation Fund Denmark, the Villum Foundation, the Danish National Research Foundation and the ARC via the Centre of Excellence in Engineered Quantum Systems (EQuS), project number CE170100009. Work at Purdue was supported by the U.S. Department of Energy, Office of Basic Energy Sciences, Division of Materials Sciences and Engineering under Award No. DE-SC0006671. Additional support from Nokia Bell Labs for the GaAs MBE effort is also gratefully acknowledged.

\bibliographystyle{naturemag}
\bibliography{bibliography,temp_bib}{}

\begin{thebibliography}{10}
\expandafter\ifx\csname url\endcsname\relax
  \def\url#1{\texttt{#1}}\fi
\expandafter\ifx\csname urlprefix\endcsname\relax\def\urlprefix{URL }\fi
\providecommand{\bibinfo}[2]{#2}
\providecommand{\eprint}[2][]{\url{#2}}

\bibitem{Cronenwett1998}
\bibinfo{author}{Cronenwett, S.~M.}, \bibinfo{author}{Oosterkamp, T.~H.} \&
  \bibinfo{author}{Kouwenhoven, L.~P.}
\newblock \bibinfo{title}{{A Tunable Kondo Effect in Quantum Dots}}.
\newblock \emph{\bibinfo{journal}{Science}} \textbf{\bibinfo{volume}{281}},
  \bibinfo{pages}{540--544} (\bibinfo{year}{1998}).

\bibitem{Potok2007}
\bibinfo{author}{Potok, R.~M.}, \bibinfo{author}{Rau, I.~G.},
  \bibinfo{author}{Shtrikman, H.}, \bibinfo{author}{Oreg, Y.} \&
  \bibinfo{author}{Goldhaber-Gordon, D.}
\newblock \bibinfo{title}{{A Tunable Kondo Effect in Quantum Dots}}.
\newblock \emph{\bibinfo{journal}{Nature}} \textbf{\bibinfo{volume}{446}},
  \bibinfo{pages}{167--171} (\bibinfo{year}{2007}).

\bibitem{Craig2004}
\bibinfo{author}{Craig, N.~J.}
\newblock \bibinfo{title}{{Tunable Nonlocal Spin Control in a Coupled-Quantum
  Dot System}}.
\newblock \emph{\bibinfo{journal}{Science}} \textbf{\bibinfo{volume}{304}},
  \bibinfo{pages}{565--567} (\bibinfo{year}{2004}).

\bibitem{Malinowski2018}
\bibinfo{author}{Malinowski, F.~K.} \emph{et~al.}
\newblock \bibinfo{title}{{Spin of a Multielectron Quantum Dot and Its
  Interaction with a Neighboring Electron}}.
\newblock \emph{\bibinfo{journal}{Physical Review X}}
  \textbf{\bibinfo{volume}{8}}, \bibinfo{pages}{011045} (\bibinfo{year}{2018}).

\bibitem{Petta2005}
\bibinfo{author}{Petta, J.~R.} \emph{et~al.}
\newblock \bibinfo{title}{{Coherent Manipulation of Coupled Electron Spins in
  Semiconductor Quantum Dots}}.
\newblock \emph{\bibinfo{journal}{Science}} \textbf{\bibinfo{volume}{309}},
  \bibinfo{pages}{2180--2184} (\bibinfo{year}{2005}).

\bibitem{Watson2018}
\bibinfo{author}{Watson, T.~F.} \emph{et~al.}
\newblock \bibinfo{title}{{A programmable two-qubit quantum processor in
  silicon}}.
\newblock \emph{\bibinfo{journal}{Nature}} \textbf{\bibinfo{volume}{555}},
  \bibinfo{pages}{633--637} (\bibinfo{year}{2018}).

\bibitem{Zajac2018}
\bibinfo{author}{Zajac, D.~M.} \emph{et~al.}
\newblock \bibinfo{title}{{Resonantly driven CNOT gate for electron spins}}.
\newblock \emph{\bibinfo{journal}{Science}} \textbf{\bibinfo{volume}{359}},
  \bibinfo{pages}{439--442} (\bibinfo{year}{2018}).

\bibitem{Nichol2017}
\bibinfo{author}{Nichol, J.~M.} \emph{et~al.}
\newblock \bibinfo{title}{{High-fidelity entangling gate for double-quantum-dot
  spin qubits}}.
\newblock \emph{\bibinfo{journal}{npj Quantum Information}}
  \textbf{\bibinfo{volume}{3}}, \bibinfo{pages}{3} (\bibinfo{year}{2017}).

\bibitem{Samkharadze2018}
\bibinfo{author}{Samkharadze, N.} \emph{et~al.}
\newblock \bibinfo{title}{{Strong spin-photon coupling in silicon.}}
\newblock \emph{\bibinfo{journal}{Science}} \textbf{\bibinfo{volume}{359}},
  \bibinfo{pages}{1123--1127} (\bibinfo{year}{2018}).

\bibitem{Mi2018}
\bibinfo{author}{Mi, X.} \emph{et~al.}
\newblock \bibinfo{title}{{A Coherent Spin-Photon Interface in Silicon}}.
\newblock \emph{\bibinfo{journal}{Nature}} \textbf{\bibinfo{volume}{555}},
  \bibinfo{pages}{559--603} (\bibinfo{year}{2018}).

\bibitem{Landig2017}
\bibinfo{author}{Landig, A.~J.} \emph{et~al.}
\newblock \bibinfo{title}{{Coherent spin-qubit photon coupling}}.
\newblock \emph{\bibinfo{journal}{arXiv}} \bibinfo{pages}{arXiv: 1711.01932}
  (\bibinfo{year}{2017}).

\bibitem{Taylor2007}
\bibinfo{author}{Taylor, J.} \emph{et~al.}
\newblock \bibinfo{title}{{Relaxation, dephasing, and quantum control of
  electron spins in double quantum dots}}.
\newblock \emph{\bibinfo{journal}{Physical Review B}}
  \textbf{\bibinfo{volume}{76}}, \bibinfo{pages}{035315}
  (\bibinfo{year}{2007}).

\bibitem{Dial2013}
\bibinfo{author}{Dial, O.~E.} \emph{et~al.}
\newblock \bibinfo{title}{{Charge Noise Spectroscopy Using Coherent Exchange
  Oscillations in a Singlet-Triplet Qubit}}.
\newblock \emph{\bibinfo{journal}{Physical Review Letters}}
  \textbf{\bibinfo{volume}{110}}, \bibinfo{pages}{146804}
  (\bibinfo{year}{2013}).

\bibitem{Taylor2005a}
\bibinfo{author}{Taylor, J.~M.} \emph{et~al.}
\newblock \bibinfo{title}{{Fault-tolerant architecture for quantum computation
  using electrically controlled semiconductor spins}}.
\newblock \emph{\bibinfo{journal}{Nature Physics}}
  \textbf{\bibinfo{volume}{1}}, \bibinfo{pages}{177--183}
  (\bibinfo{year}{2005}).

\bibitem{Burkard2006}
\bibinfo{author}{Burkard, G.} \& \bibinfo{author}{Imamoglu, A.}
\newblock \bibinfo{title}{{Ultra-long-distance interaction between spin
  qubits}}.
\newblock \emph{\bibinfo{journal}{Physical Review B}}
  \textbf{\bibinfo{volume}{74}}, \bibinfo{pages}{041307(R)}
  (\bibinfo{year}{2006}).

\bibitem{Baart2017}
\bibinfo{author}{Baart, T.~A.}, \bibinfo{author}{Fujita, T.},
  \bibinfo{author}{Reichl, C.}, \bibinfo{author}{Wegscheider, W.} \&
  \bibinfo{author}{Vandersypen, L. M.~K.}
\newblock \bibinfo{title}{{Coherent spin-exchange via a quantum mediator}}.
\newblock \emph{\bibinfo{journal}{Nature Nanotechnology}}
  \textbf{\bibinfo{volume}{12}}, \bibinfo{pages}{26--30}
  (\bibinfo{year}{2016}).

\bibitem{Mehl2014a}
\bibinfo{author}{Mehl, S.}, \bibinfo{author}{Bluhm, H.} \&
  \bibinfo{author}{DiVincenzo, D.~P.}
\newblock \bibinfo{title}{{Two-qubit couplings of singlet-triplet qubits
  mediated by one quantum state}}.
\newblock \emph{\bibinfo{journal}{Physical Review B}}
  \textbf{\bibinfo{volume}{90}}, \bibinfo{pages}{045404}
  (\bibinfo{year}{2014}).

\bibitem{Srinivasa2015}
\bibinfo{author}{Srinivasa, V.}, \bibinfo{author}{Xu, H.} \&
  \bibinfo{author}{Taylor, J.}
\newblock \bibinfo{title}{{Tunable Spin-Qubit Coupling Mediated by a
  Multielectron Quantum Dot}}.
\newblock \emph{\bibinfo{journal}{Physical Review Letters}}
  \textbf{\bibinfo{volume}{114}}, \bibinfo{pages}{226803}
  (\bibinfo{year}{2015}).

\bibitem{Croot2017}
\bibinfo{author}{Croot, X.~G.} \emph{et~al.}
\newblock \bibinfo{title}{{Device Architecture for Coupling Spin Qubits Via an
  Intermediate Quantum State}}.
\newblock \emph{\bibinfo{journal}{arXiv}} \bibinfo{pages}{arXiv: 1707.06479}
  (\bibinfo{year}{2017}).

\bibitem{Martins2016}
\bibinfo{author}{Martins, F.} \emph{et~al.}
\newblock \bibinfo{title}{{Noise suppression using symmetric exchange gates in
  spin qubits}}.
\newblock \emph{\bibinfo{journal}{Physical Review Letters}}
  \textbf{\bibinfo{volume}{116}}, \bibinfo{pages}{116801}
  (\bibinfo{year}{2016}).

\bibitem{Reed2016}
\bibinfo{author}{Reed, M.~D.} \emph{et~al.}
\newblock \bibinfo{title}{{Reduced sensitivity to charge noise in semiconductor
  spin qubits via symmetric operation}}.
\newblock \emph{\bibinfo{journal}{Physical Review Letters}}
  \textbf{\bibinfo{volume}{116}}, \bibinfo{pages}{110402}
  (\bibinfo{year}{2016}).

\bibitem{Martins2017}
\bibinfo{author}{Martins, F.} \emph{et~al.}
\newblock \bibinfo{title}{{Negative Spin Exchange in a Multielectron Quantum
  Dot}}.
\newblock \emph{\bibinfo{journal}{Physical Review Letters}}
  \textbf{\bibinfo{volume}{119}}, \bibinfo{pages}{227701}
  (\bibinfo{year}{2017}).

\bibitem{Deng2017}
\bibinfo{author}{Deng, K.}, \bibinfo{author}{Mayhall, N.~J.} \&
  \bibinfo{author}{Barnes, E.}
\newblock \bibinfo{title}{{Negative exchange interactions in coupled
  few-electron quantum dots}}.
\newblock \emph{\bibinfo{journal}{arXiv}} \bibinfo{pages}{arXiv: 1712.05795}
  (\bibinfo{year}{2017}).

\bibitem{Malinowski2017a}
\bibinfo{author}{Malinowski, F.~K.} \emph{et~al.}
\newblock \bibinfo{title}{{Spectrum of the Nuclear Environment for GaAs Spin
  Qubits}}.
\newblock \emph{\bibinfo{journal}{Physical Review Letters}}
  \textbf{\bibinfo{volume}{118}}, \bibinfo{pages}{177702}
  (\bibinfo{year}{2017}).

\bibitem{Kurland2000}
\bibinfo{author}{Kurland, I.~L.}, \bibinfo{author}{Aleiner, I.~L.} \&
  \bibinfo{author}{Altshuler, B.~L.}
\newblock \bibinfo{title}{{Mesoscopic magnetization fluctuations for metallic
  grains close to the Stoner instability}}.
\newblock \emph{\bibinfo{journal}{Physical Review B}}
  \textbf{\bibinfo{volume}{62}}, \bibinfo{pages}{14886--14897}
  (\bibinfo{year}{2000}).

\bibitem{Folk2001}
\bibinfo{author}{Folk, J.~A.} \emph{et~al.}
\newblock \bibinfo{title}{{Ground State Spin and Coulomb Blockade Peak Motion
  in Chaotic Quantum Dots}}.
\newblock \emph{\bibinfo{journal}{Physica Scripta}}
  \textbf{\bibinfo{volume}{T90}}, \bibinfo{pages}{26--33}
  (\bibinfo{year}{2001}).

\bibitem{Medford2013}
\bibinfo{author}{Medford, J.} \emph{et~al.}
\newblock \bibinfo{title}{{Self-consistent measurement and state tomography of
  an exchange-only spin qubit}}.
\newblock \emph{\bibinfo{journal}{Nature Nanotechnology}}
  \textbf{\bibinfo{volume}{8}}, \bibinfo{pages}{654--659}
  (\bibinfo{year}{2013}).

\bibitem{Kim2014}
\bibinfo{author}{Kim, D.} \emph{et~al.}
\newblock \bibinfo{title}{{Quantum control and process tomography of a
  semiconductor quantum dot hybrid qubit}}.
\newblock \emph{\bibinfo{journal}{Nature}} \textbf{\bibinfo{volume}{511}},
  \bibinfo{pages}{70} (\bibinfo{year}{2014}).

\bibitem{Cao2016}
\bibinfo{author}{Cao, G.} \emph{et~al.}
\newblock \bibinfo{title}{{A Tunable Hybrid Qubit in a GaAs Double Quantum
  Dot}}.
\newblock \emph{\bibinfo{journal}{Physical Review Letters}}
  \textbf{\bibinfo{volume}{116}}, \bibinfo{pages}{086801}
  (\bibinfo{year}{2016}).

\bibitem{Fujita2017}
\bibinfo{author}{Fujita, T.}, \bibinfo{author}{Baart, T.~A.},
  \bibinfo{author}{Reichl, C.}, \bibinfo{author}{Wegscheider, W.} \&
  \bibinfo{author}{Vandersypen, L. M.~K.}
\newblock \bibinfo{title}{{Coherent shuttle of electron-spin states}}.
\newblock \emph{\bibinfo{journal}{npj Quantum Information}}
  \textbf{\bibinfo{volume}{3}}, \bibinfo{pages}{22} (\bibinfo{year}{2017}).

\bibitem{Maune2012}
\bibinfo{author}{Maune, B.~M.} \emph{et~al.}
\newblock \bibinfo{title}{{Coherent singlet-triplet oscillations in a
  silicon-based double quantum dot}}.
\newblock \emph{\bibinfo{journal}{Nature}} \textbf{\bibinfo{volume}{481}},
  \bibinfo{pages}{344--347} (\bibinfo{year}{2012}).

\bibitem{Flentje2017}
\bibinfo{author}{Flentje, H.} \emph{et~al.}
\newblock \bibinfo{title}{{Coherent long-distance displacement of individual
  electron spins}}.
\newblock \emph{\bibinfo{journal}{Nature Communications}}
  \textbf{\bibinfo{volume}{8}}, \bibinfo{pages}{501} (\bibinfo{year}{2017}).

\bibitem{Gaudreau2011}
\bibinfo{author}{Gaudreau, L.} \emph{et~al.}
\newblock \bibinfo{title}{{Coherent control of three-spin states in a triple
  quantum dot}}.
\newblock \emph{\bibinfo{journal}{Nature Physics}}
  \textbf{\bibinfo{volume}{8}}, \bibinfo{pages}{54--58} (\bibinfo{year}{2011}).

\end{thebibliography}

%Supplementary Materials

\newpage

\setcounter{figure}{0}
\setcounter{equation}{0}
\renewcommand{\thefigure}{S\arabic{figure}}  
\renewcommand{\theequation}{S\arabic{equation}}
\renewcommand{\thetable}{S\arabic{table}}

\section*{METHODS}

\section{Dot preparation \& readout}
The array of quantum dots is defined in a high-mobility ($230$ m$^2$/Vs) two-dimensional electron gas (density $2.5 \times 10^{15}$ m$^{-2}$) located 57 nm below the surface of a GaAs/AlGaAs heterostructure, by means of electrostatic gate electrodes deposited on top of the heterostructure~\cite{Malinowski2018}.  A layer of HfO2 with 10 nm thickness is deposited on top of the heterostructure, prior to patterning the gold electrodes by electron-beam and lift-off lithography.

Within each double dot, spin-to-charge conversion is used to read out the relative spin alignment within each double dot\cite{Petta2005}. Specifically, a frequency-multiplexed measurement pulse reflected off two proximal radio-frequency quantum-dot-based charge sensors allows us to distinguish between singlet and triplet states of each double dot, independently and with single-shot fidelity.

\section{Determination of gate-voltage pulses}

The linear geometry of the quintuple quantum dot makes it difficult to measure a five-dimensional five-dot charge stability diagram: the central dot can exchange electrons with the reservoirs only via (co-)tunneling through the left or right double dot, which is strongly suppressed, particularly once the device is tuned up. 
Instead of mapping out full charge stability diagrams in order to determine pulse amplitudes and pulse directions in gate-voltage space, we proceed in steps. First, we choose  the readout voltages $V_{j}^R$ (which we refer to as readout point), then the separation voltages $V_{j}^S$ (referred to as separation point), and finally interaction voltages $V_{j}^I$ (referred to as interaction point). Here, $j \in \{L1, L2, M, R1, R2\}$, see Fig.~\ref{fig1}a. 

For each double dot (i.e. left and right double dot separately) we establish a partial charge stability diagram, by sweeping its plunger gates ($V_{L1,L2}$ or $V_{R1,R2}$) while monitoring its charge sensor. Application of (unoptimized) pulse sequences (corresponding to double-dot leakage-spectroscopy measurement at fixed, finite magnetic field) allows us to optimize the static gate voltages associated with each double dot and charge sensor to obtain suitable single-shot readout performance. 
These readout voltages $V_{j}^R$ define the origin of our coordinate system. With respect to $V_{j}^R$ we then define detuning parameters $\varepsilon_L = (V_{L2}-V_{L1})/\sqrt{2}$ and $\varepsilon_R = (V_{R1}-V_{R2})/\sqrt{2}$ for each double dot. 

Having defined $\varepsilon_{L/R}$ we repeat leakage spectroscopy, a generalization of ``spin funnel'' measurements~\cite{Petta2005,Maune2012}. For the case of one double dot coupled to a multielectron dot, this procedure is described in detail in Refs.~\onlinecite{Martins2017,Malinowski2018}.
For this device, we apply leakage spectroscopy pulses to both double dots simultaneously, while varying values of $\varepsilon_L$, $\varepsilon_R$, and the applied magnetic field $B$. 
This yields data as in the left-most regions of Fig.~\ref{figS1}a,b. Phenomenologically, the flattening of the curved leakage feature towards increasing $\varepsilon_{L,R}$ informs us about the decreasing strength of the residual exchange coupling within each double dot. This allows us to choose the separation point of each double dot $V_j^S$, by choosing $\varepsilon_{L,R}$ such that the leakage feature lies between $B=10$ and $20$~mT. 
For instance, in Fig.~\ref{figS1} we chose $\varepsilon_L^S=13$~mV and $\varepsilon_R^S=18$~mV as the separation point.

The separation point $V_{j}^S$ in turn serves as a reference point for determining the interaction point $V_{j}^I$, at which the multielectron-dot-mediated exchange interaction is induced. In case of the data presented in Figs. 2b,c and \ref{figS1}a,b the interaction point $V_j^I$ is parametrized by $\DP$ according to the formula:

\begin{equation}
	\left(
		\begin{array}{c}
			V_{L1}^I \\
			V_{L2}^I \\
			V_{M}^I \\
			V_{R1}^I \\
			V_{R2}^I
		\end{array}
	\right) 
	= %
	\left(
		\begin{array}{c}
			V_{L1}^S \\
			V_{L2}^S \\
			V_{M}^S \\
			V_{R1}^S \\
			V_{R2}^S
		\end{array}
	\right)
	+ \frac{\DP}{\sqrt{35}}
	\left(
		\begin{array}{c}
			-3 \\
			-2 \\
			3 \\
			-2 \\
			-3
		\end{array} 
	\right)
	\label{parameters1}
\end{equation}
Equation \ref{parameters1} implements negative voltage pulses on plunger gates of the two double dots, with the intention of suppressing exchange of electrons between the quintuple dot and the reservoirs during the (positive) interaction pulse on $V_M$.
With this choice, the ``cross-compensation'' pulses applied to $V_{L1,L2, R1, R2}$ are \emph{proportional} in amplitude to $\DP$,  which distinguishes these pulses from $\MP$ pulses (which employ \emph{constant} cross-compensation pulses, as described below). 
The normalization factor $\sqrt{35}$ ensures that a change of $\DP$ by 1~mV corresponds to a distance of 1~mV in the gate voltage space with a Cartesian metric. 

We found that it is not necessary to fine tune the amplitudes of the compensating pulses, presumably due to the effective isolation of the multielectron dot from the reservoirs~\cite{Flentje2017}. This simplifies the generation of subnanosecond pulses, as explained in Methods Section~III. Therefore, for some data sets we only vary $V_{M}^I$, $V_{L2}^I$, and $V_{R1}^I$. In these cases, we parametrize the interaction point $V_j^I$ by parameters $\MP$ and $\varepsilon$:
\begin{multline}
	\left(
		\begin{array}{c}
			V_{L1}^I \\
			V_{L2}^I \\
			V_{M}^I \\
			V_{R1}^I \\
			V_{R2}^I
		\end{array}
	\right) 
	= 
	\left(
		\begin{array}{c}
			V_{L1}^S \\
			V_{L2}^S \\
			V_{M}^R \\ 
			V_{R1}^S \\
			V_{R2}^S
		\end{array}
	\right)
	+ 
	\left(
		\begin{array}{c}
			0 \\
			0 \\
			\MP \\
			0 \\
			0 \\
		\end{array}
	\right)
	+ \\
	+ \frac{\varepsilon}{\sqrt{2}} 
	\left(
		\begin{array}{c}
			0 \\
			-1 \\
			0 \\
			+1 \\
			0
		\end{array}
	\right)
	+ \frac{\varepsilon_X}{\sqrt{35}}
	\left(
		\begin{array}{c}
			-3 \\
			-2 \\
			0 \\
			-2 \\
			-3
		\end{array} 
	\right).
	\label{MPpulses}
\end{multline}
This parametrization enables us to fine-tune the interaction time with subnanosecond resolution by varying only $\MP$, thereby needing only one additional channel of the arbitrary waveform generator (Methods Section~III).
Physically, the parameter $\varepsilon$ controls the relative detuning between the chemical potential of the left and right inner dot. 
The parameter $\varepsilon_X$ (which we keep fixed) implements cross-compensation amplitudes that are independent of $\MP$.

Aside from this technical difference between pulses parametrized by $\DP$ (Eq.~\ref{parameters1}) and $\MP$ (Eq.~\ref{MPpulses}), data presented in this article was acquired using slightly different voltages applied to the static gate electrodes, as well as different choices of $V_j^S$ and $\varepsilon_X$ (Table~\ref{tunings}). However, the general principle of tuning up pulse sequences in either case was similar to the pocedure described here. No significant retuning of the quintuple-dot array was necessary in between different data sets, and therefore the tunnel couplings can be considered unchanged throughout the entire experiment. However, precise choices of measurement points ($V_j^R$), separation points ($V_j^S$), sensor settings, as well as $\varepsilon_X$ were adjusted between data sets.

\begin{table}[tb]
\caption{
	Static tuning configurations used in different measurements.
	}
\label{tunings}
\begin{tabular}{c|c}
\textbf{DC configuration} & \textbf{Figures} \\ \hline
1 & \ref{fig2}a, \ref{fig3}b \ref{fig4}b, \ref{figS3} \\ 
2 & \ref{fig4}a \\
3 & \ref{fig2}b,c, \ref{figS2} \\ 
4 & \ref{figS1} \\
\end{tabular} 
\end{table}

\section{Implementation of exchange pulses with subnanosecond resolution}

To achieve subnanosecond resolution of the exchange pulse we interfered two nominally cancelling signals generated by two arbitrary waveform generator channels, and applied the combined signal to the multielectron-dot plunger gate $V_M$. Specifically, we set the two channels to output a square waveform of identical duration and amplitude, but with opposite polarity, and combine them using an inverted power splitter. The pulse period is set to the repetition time of the intended pulse sequence, the rising edge of the pulse is set to the beginning of the intended exchange pulse, while the falling edge happens at the beginning of the double-dot initialization step. By finely adjusting the channel skew of the arbitrary waveform generator, positive or negative $V_M$ pulses can be generated with subnanosecond control. While this method allows to overcome the limitations of the waveform generator's temporal resolution of 1.2 GS/s (Tektronix AWG 5014C), the effective voltage pulse reaching the gate electrodes is still constrained by the 0.8~ns pulse rise time in our dilution refrigerator, resulting in distortion effects in Figs.~\ref{fig2}, \ref{fig3} and \ref{fig4} (cf. Methods Section~VIII).

\section{Level structure of the multielectron dot inferred from spin-leakage spectroscopy}

If two spin states with different total spin projection $\hat{S}_z$ are brought together in energy for a sufficiently long time, leakage from one state to the other can occur due to higher-order (non-spin-conserving) elastic processes. This provides an experimental method, spin leakage spectroscopy~\cite{Malinowski2018}, to experimentally detect discrete states, and to quantify the exchange interaction by comparison with the Zeeman energy. 
In the conventional case of a double dot it can be used to locate the position of the crossing between the singlet $\ket{S}$ and the fully polarized triplet $\ket{T_{+/-}}$ state (the sign of the electronic g-factor defines which of the triplet states is used), and results in the characteristic funnel shape~\cite{Petta2005,Maune2012}. In the case of a triple quantum dot the position of an analogous crossing, which depends on the value of the external magnetic field, enables the reconstruction of the exchange profile~\cite{Gaudreau2011,Martins2017,Malinowski2018}. Here we employ the same technique to the case of the two double quantum dots coupled to the multielectron dot.

\begin{figure*}
	\centering
	\includegraphics[width=\textwidth]{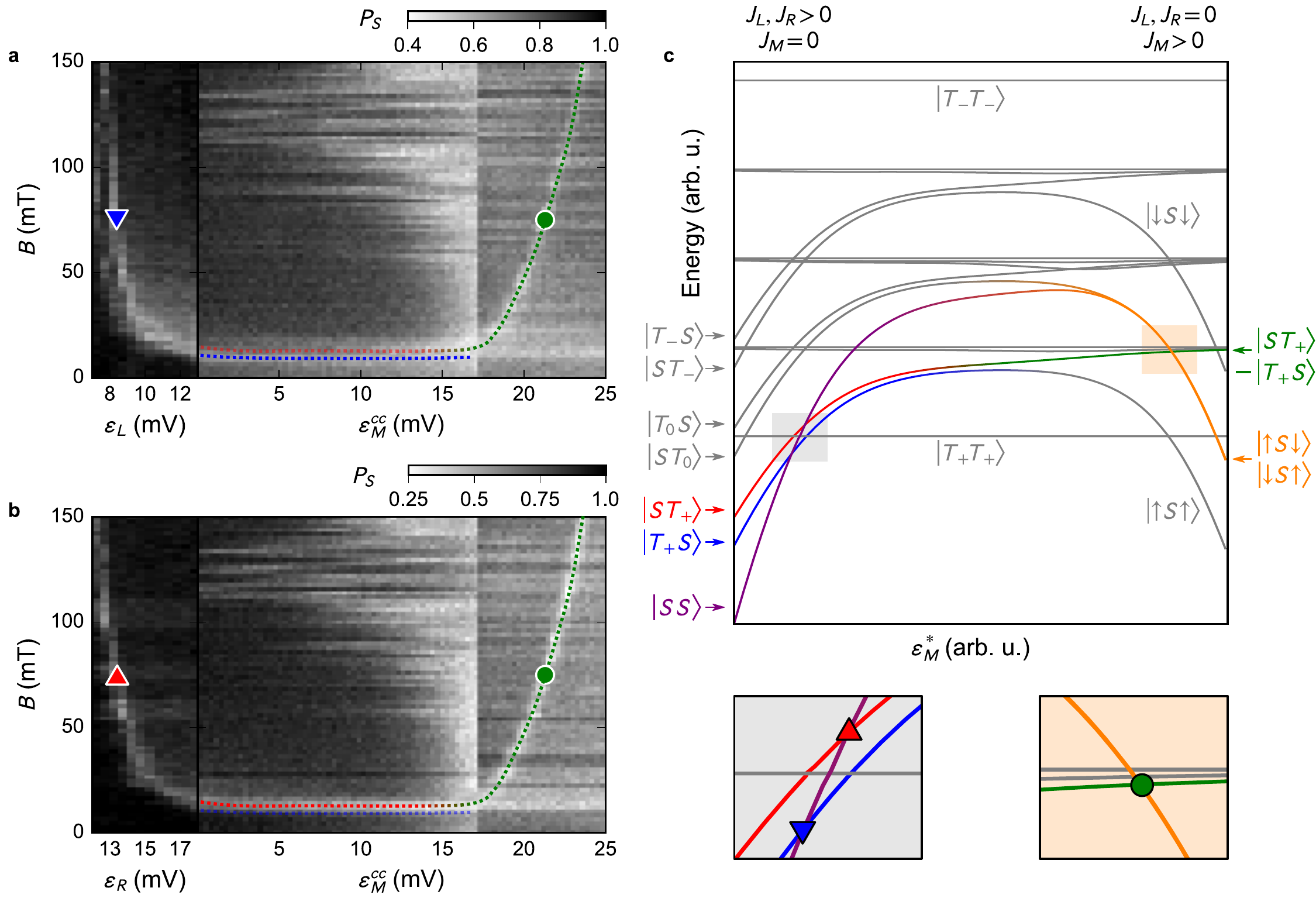} 
	\caption{
	{\bf Leakage spectroscopy using two exchange-coupled double quantum dots.}
	Leakage spectroscopy measurement performed simultaneously for the left ({\bf a}) and the right ({\bf b}) double quantum dot.
	Data along $\DP$ was acquired as two separate data sets with different sweep time between reference and separation points, resulting in a vertical artifact at $\DP=17$~mV. The bright horizontal feature at $B\approx15$~mT for $\DP>17$~mV is a result of leakage during the separation step of the applied pulse sequence (cf. Methods Section~II) and does not indicate an additional spin-state crossing.
	{\bf c} Schematic energy diagram of the two exchange-coupled double quantum dots, for finite in-plane magnetic field. In the left only the exchange interaction within the left and right double quantum dot ($J_{L/R}$) is non-zero. In the right only exchange mediated my the multielectron quantum dot ($J_M$) is non-zero. The multielectron dot plays the role of a barrier between the double dots (see text), and hence the kets denote the spin state of the left and right double dot only. Markers indicate the crossings detected by leakage spectroscopy measurements.
	}
	\label{figS1}
\end{figure*}

The sequence of the applied voltage pulses is the same as the one used to detect exchange oscillations mediated by the multielectron quantum dot, except that the interaction time is increased to $\tau = 150$~ns. This time is sufficiently long to allow leakage from the initialized state into other states, for those pixels for which a level crossing occurs at the interaction point. (Moreover, 150~ns is sufficiently long to wash out any remaining coherent exchange oscillations, thanks to dephasing from Overhauser field fluctuations and  charge noise.) Such pixels therefore show a suppression of $P_S$. For example, the left panels of Fig.~\ref{figS1}a,b present the position of the $S$-$T_+$ crossing for the two double quantum dots, acquired simultaneously. Each one can be viewed as the conventional ``spin funnel'' of a singlet-triplet qubit~\cite{Petta2005,Maune2012}. 
Note that the horizontal axes in Fig.~\ref{figS1}a and Fig.~\ref{figS1}b correspond to different gate-voltage parameters, namely detuning within the left ($\varepsilon_L$) and the right ($\varepsilon_R$) double quantum dot. Therefore the apparent similarity between the two funnels does not imply any interactions across the multielectron dot; it merely indicates that the intradot exchange coupling within the left and right double quantum dot had been tuned up with similar strenghts. 

%:fig1

The right panels of Fig.~\ref{figS1}a,b present the result in the regime where long-range exchange across the multielectron dot turns on. In this part of the panels the horizontal axis is shared, and denotes the pulse amplitude $\DP$. We observe that for intermediate values of $\DP$, the leakage features detected from the left (blue dotted lines) occur at different magnetic field values compared to the leakage features detected from the right (red dotted line). The associated level crossings therefore belong to different states. In contrast, for high values of $\DP$, the leakage feature detected from the left and right occur at exactly the same magnetic field values (green dotted line), and diverge towards increasing field.

This agrees with a simplistic Heisenberg model of four exchange-coupled spin-1/2 dots arranged in a linear array (i.e. the multielectron dot is simply treated as a tunnel barrier). The associated energy diagram, arising from an appropriate choice of the three pairwise exchange interactions within the array, allows us to identify the observed features (Fig.~\ref{figS1}c). In the left side of the diagram, only exchange coupling within each double quantum dot is nonzero (and the associated spin states can be written as product states between states on the left double dot, and states on the right double dot). In the right side of Fig.~\ref{figS1}c, however, it is the exchange interaction mediated by the multielectron quantum dot that becomes nonzero (and associated spin states can no longer be written as product states between left and right).

The leakage features in the left part of Fig.~\ref{figS1}a,b correspond to the $S$-$T_+$ crossing of the left and the right double dot. In the four-dot notation, these are $\ket{SS}$-$\ket{T_+ S}$ and $\ket{SS}$-$\ket{S T_+}$ crossings, which are indicated by a blue and a red triangle, respectively. In the middle part of the energy diagram and leakage spectroscopy data, $\ket{T_+ S}$ and $\ket{S T_+}$ states start to hybridize due to exchange mediated by the multielectron dot. As a result, only one of the leakage features continues (the one indicated with the red-to-green dotted line), while the other leakage feature stops (blue dotted line).
This indicates the position at which $\ket{T_+ S}$ and $\ket{S T_+}$ are no longer eigenstates, but their superposition $(\ket{T_+ S} - \ket{S T_+})/\sqrt{2}$ is (indicated by a green line in \ref{figS1}b). At this position, the $\ket{SS}$ state is also no longer an eigenstate, but instead $\ket{\uparrow S \downarrow}$ and $\ket{\downarrow S \uparrow}$ are (indicated by the orange lines).

%:fig2
\begin{figure*}
	\centering
	\includegraphics[width=\textwidth]{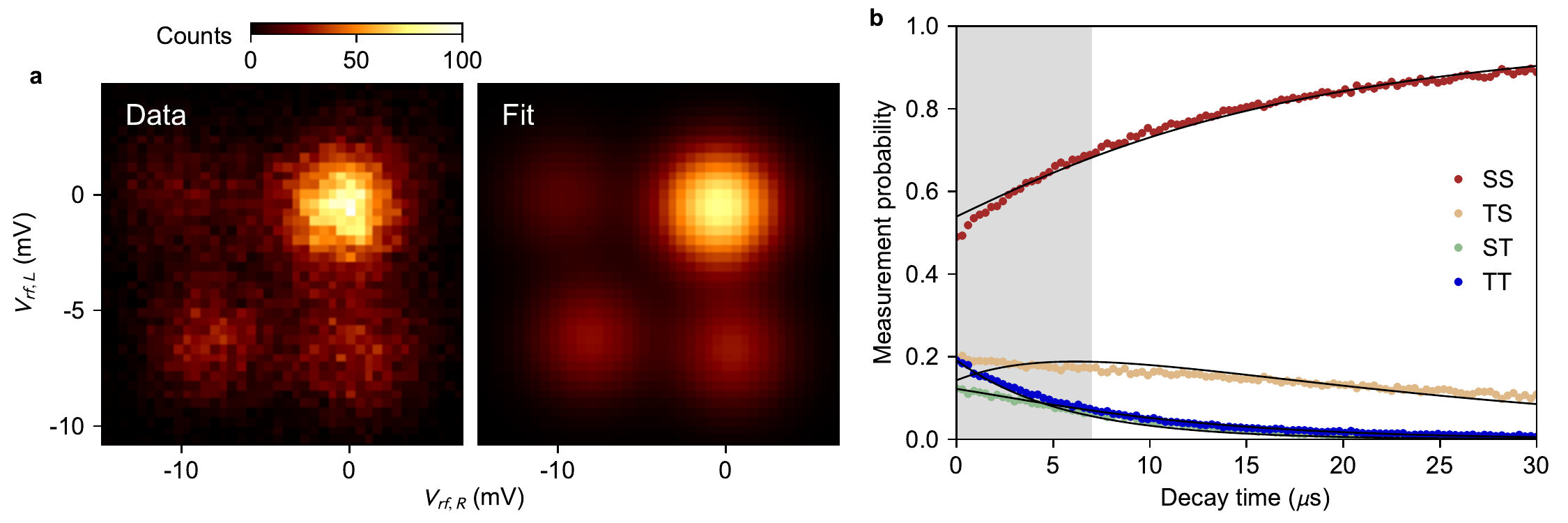}
	\caption{
	{\bf Probability estimation.}
	{\bf a} Two-dimensional histogram of measured single-shot readouts (left) and a fitted quadrupole Gaussian (right).
	{\bf b} Decay of the triplet states in the measurement configuration. The experimentally measured decay (dots) is fitted by a simple model based on two independent decay rates for the two double quantum dots (lines).
	}
	\label{figS2}
\end{figure*}

\section{Extraction of joint probabilities from histograms}

The joint probabilities, presented in Fig.~\ref{fig2}b, are calculated based on histograms of single-shot outcomes of the demodulated sensor voltages for each pulse amplitude (presented in the Supplementary Video 1). For calibration purposes, we first sum multiple histograms associated with different pulse amplitudes, in order to get sufficient counts for all four outcomes SS, ST, TS, TT.   To this two-dimensional histogram, we  fit a two-dimensional quadruple Gaussian to obtain the position of the four peaks (8 parameters) and their widths (2 parameters, one of which sets the widths for the sensor signals of the left double quantum dot, and the other sets the widths of the sensor signals of the right double quantum dot). The data and the fit are presented in Fig.~\ref{figS2}a. Having fixed the positions and widths associated with all four Gaussians, we leave their amplitudes as free fit parameters when fitting histograms separately for each voltage pulse amplitude. 
The normalized amplitudes of the Gaussians yield the measured joint probabilities $\vec{p}_\mathrm{meas}=(p_\mathrm{SS}, p_\mathrm{TS}, p_\mathrm{ST}, p_\mathrm{TT})$, uncorrected for charge relaxation of the two-electron double-dot states during the measurement interval.

To correct for the decay of the two-electron states in the left and right double dot during the measurement interval, we fix the amplitude of the exchange-inducing pulse at a value that yields significant number of counts for all four possible outcomes, and introduce a waiting time in the readout configuration before performing measurement of the sensor signals. 
This provides a measurement of the relaxation time, as exemplified in Fig.~\ref{figS2}b. We fit the data assuming independent relaxation rates, different for the two double quantum dots. This model yields:
\begin{widetext}
\begin{equation}
	\vec{p}(t) = M(t) \vec{p}(0) =
	\left(
	\begin{array}{cccc}
			1 & 1-e^{-\Gamma_L t} & 1-e^{-\Gamma_R t} & 1-e^{-\Gamma_L t}-e^{-\Gamma_R t}+e^{-(\Gamma_L+\Gamma_R) t} \\
			0 & e^{-\Gamma_L t} & 0 & e^{-\Gamma_L t}-e^{-(\Gamma_L+\Gamma_R) t} \\
			0 & 0 & e^{-\Gamma_R t} & e^{-\Gamma_R t}-e^{-(\Gamma_L+\Gamma_R) t} \\
			0 & 0 & 0 & e^{(-\Gamma_L-\Gamma_R) t}
	\end{array}
	\right)
	\vec{p}(0)
\end{equation}
\end{widetext}
where $\Gamma_L = 0.5$~MHz and $\Gamma_R = 0.13$~MHz are relaxation rates in the left and right double quantum dot, respectively.

Having fitted the decay rates for both double dots (see Fig.~\ref{figS2}b) we can reverse the relation between measured probabilities and the actual probabilities:
\begin{equation}
	\vec{p}_\mathrm{meas} = \frac{1}{T_R} \int\limits_0^{T_R} M(t) \vec{p}_\mathrm{act} \mathrm{d}t
	\label{decay}
\end{equation}
where $\vec{p}_\mathrm{meas/act}$ are the vectors of measured/actual outcome probabilities, $M(t)$ captures the decay during the waiting time $t$ and $T_R$ is the total readout time of 7~$\mu$s (as indicated with the gray-shaded region in Fig.~\ref{figS2}b). The integration is performed to include decay that occurs \emph{during} the readout time. Application of the numerically inversed relation~\ref{decay} yields the calculated joint probabilities of the four states, reported in Fig.~\ref{fig3}c.

\section{Measurement of the charge distribution at the spin-interaction points}

%:fig3
\begin{figure}
	\centering
	\includegraphics[width=0.48\textwidth]{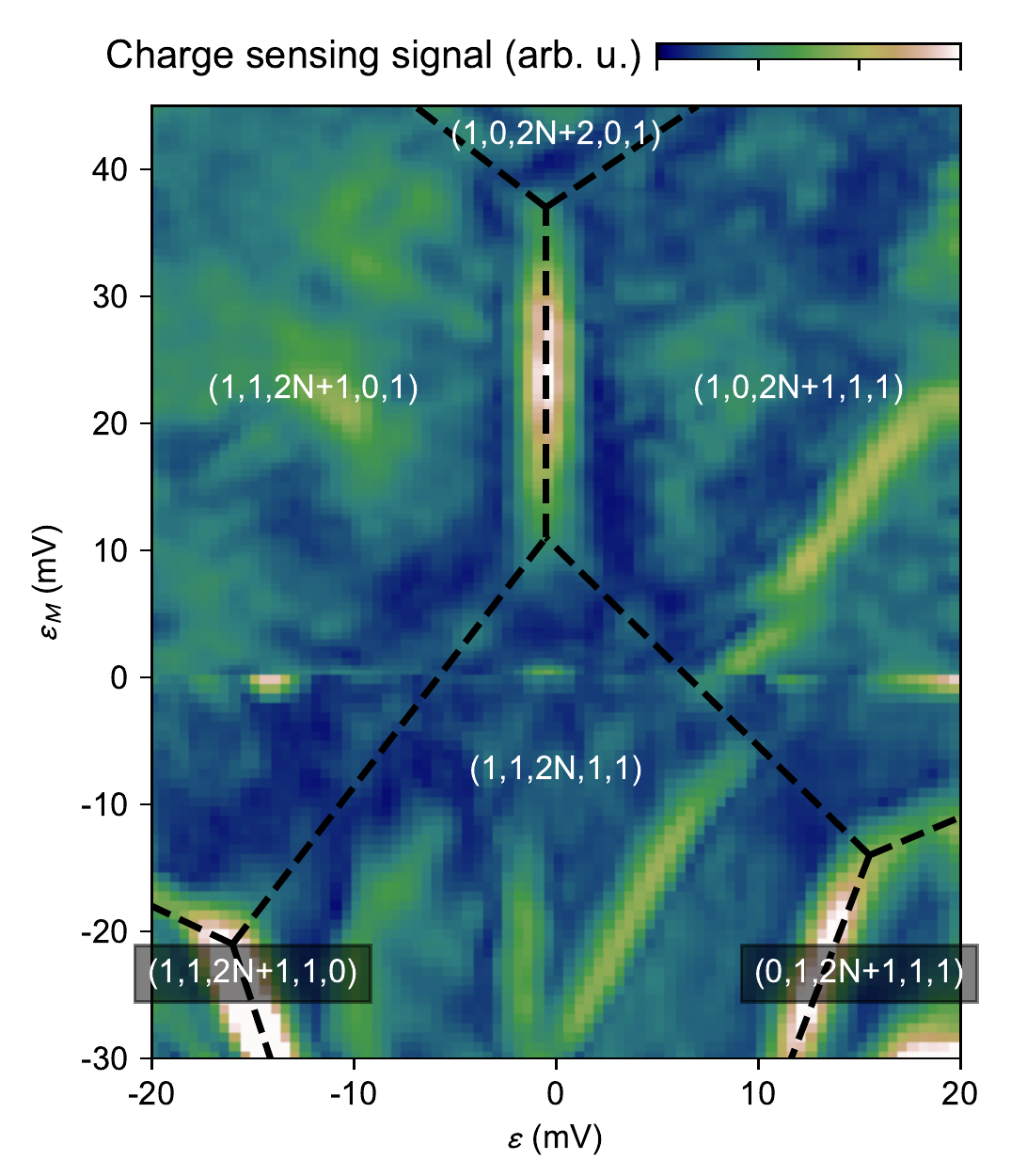}
	\caption{
	Processed diagram of the charge distribution during the interaction mediated by the multielectron quantum dot. Dashed lines indicate the extracted positions of the charge transitions.
	}
	\label{figS3}
\end{figure}

To independently confirm the position of the electrons during the interaction step we extend the interaction time to 4~$\mu$s, while maintaining the remainder of the pulse sequence unchanged (the nanosecond-scale interaction times used in Fig.~\ref{fig3} would be too short to allow the radiofrequency tank circuits to respond). During this 4~$\mu$s-long time we apply a radiofrequency measurement tone to both charge sensors, and record their (demodulated) response while varying $\varepsilon$ and $\MP$. Due to the capacitive cross-coupling between gate electrodes of the quintuple dot and the sensor quantum dots, we acquire such charge-sensing maps for several different settings of the charge sensors. This is needed because a sensor signal is sensitive to charge in the device only when the sensor dot's operating point falls on the positive or negative slope of one of its Coulomb-oscillation conductance peaks. We perform the numerical derivative of each data set along $\varepsilon$, then apply blur by convolving the result with a Gaussian kernel ($\sigma=1.5$ pixel), and take the absolute value to remove sign changes of the sensor's sensitivity when its operating point switches from the positive slope to the negative slope. Finally, we sum the obtained data sets with different weights, for best visibility of charge transitions within the device. 
The processed data obtained in this way is presented in Fig.~\ref{figS3}. The inferred charge transitions are indicated with dashed black lines. The features corresponding to the electron transfer from each of the inner dots to the multielectron dot are very weak, due to the large tunnel coupling chosen for investigating the superexchange regime. The two additional regions in the bottom left and right of Fig.~\ref{figS3} correspond to, respectively, (1,1,$2N \! + \! 1$,1,0) and (0,1,$2N \! + \! 1$,1,1) charge configurations of the quintuple quantum dot, which are partially visible also in Fig.~\ref{fig3}b. They correspond to the relocation of the outer reference spins to the inner dots, and hence we do not inspect these regions further.

In addition to the indicated charge transitions, several other strong features are observed in Fig.~\ref{figS3}, which have no counterparts in the data of exchange oscillations (Fig.~\ref{fig3}b). We associate these with artifacts arising from the long interaction time of 4~$\mu$s, which allows charge transitions within the metastable electron configuration of the quintuple dot. As long as these relaxation processes are sufficiently slow, they are irrelevant when operating exchange oscillations with short $\tau$~\cite{Flentje2017}, and possibly could be suppressed by suitable cross compensation pulses.

\section{Hubbard model of the exchange interaction}

\begin{figure}
	\centering
	\includegraphics[width=0.48\textwidth]{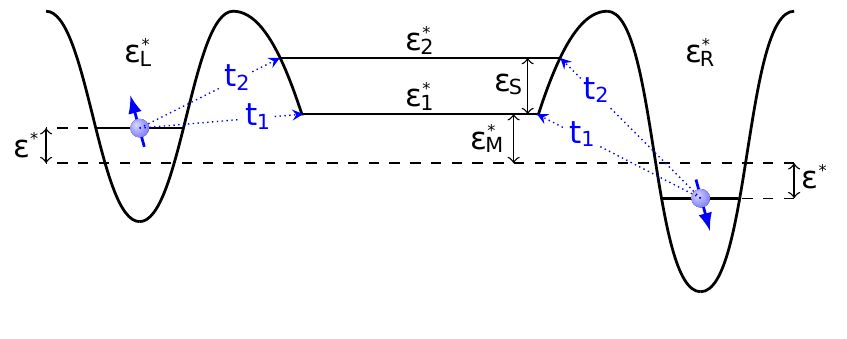}
	\caption{Schematic of a spinless multielectron dot (center) tunnel-coupled to two single-electron quantum dots (left and right). Symbols $\varepsilon_{L/1/2/R}$ label the single-particle energies of the orbitals in single-electron dots, and the two lowest unoccupied orbitals in the multielectron dot. The energy difference between the two orbitals in the multielectron dot is denoted by $\varepsilon_S = \varepsilon_2 - \varepsilon_1$. The parameters $\varepsilon^*$ and $\varepsilon_M^*$ are varied to obtain Fig.~\ref{fig3}c. Electrons in the middle dot (light blue) are singlet-paired below the Fermi energy. The electrons in left and right dots (blue) are only tunnel-coupled to unoccupied orbitals (blue dotted arrows). We assume that tunneling rates from left and right dots are equal.}
	\label{figS4}
\end{figure}

The exchange oscillation simulations, presented in Fig.~\ref{fig3}d, have been obtained using a phenomenological model for a multielectron quantum dot outlined in Ref. \onlinecite{Malinowski2018} (see also Ref.~\onlinecite{Deng2017}), by adding terms which describe two tunnel-coupled single-electron quantum dots. For simplicity we only model the three dots in the center of the quintuple-dot array. The outer dots are decoupled from the inner dots for the period of interaction and will not contribute to the effective exchange mediated by the middle dot. (Also, while most experimental control parameters are voltages, such as $\varepsilon, \MP$ etc, corresponding parameters in our model are energies, such as $\varepsilon^*, \varepsilon_M^*$ etc. Because of the negative electronic charges, an increase in $\MP$ in the experiment, for instance, corresponds in the simulation to making $\varepsilon_M^*$ more negative.) Since the multielectron dot has a spinless ground state, we neglect the electron pairs that are singlet-paired in the orbitals below the Fermi energy. We also neglect all but the two lowest unoccupied orbitals, such that the three dots are described by a Hubbard model with four orbitals. The labels $L$ and $R$ denote the orbitals of the two inner dots, and labels $1$ and $2$ correspond to the two,  non-degenerate orbitals of the middle dot. The Hubbard Hamiltonian of the system (illustrated in Fig.~\ref{figS4}) is given by
\begin{multline}
\label{eqnS2}
\hat{H} = \sum_{i}\left( \varepsilon_i^* \hat{n}_i + \frac{U_i}{2}\hat{n}_i(\hat{n}_i-1)\right) + \sum_{i\neq j}\frac{K_{ij}}{2}\hat{n}_i \hat{n}_{j} + \\
- \frac{\xi}{2}\hat{S}^2- \sum_{\langle i,j \rangle}\sum_\alpha t_{ij}(c^\dagger_{i,\alpha}c_{j,\alpha}+\mathrm{H.c.}),
\end{multline}
which sums over the orbitals $i=L,1,2,R$ and electron spin orientations $\alpha= \uparrow,\downarrow$. The operator $\hat{n}_i=\sum_{\alpha}c^\dagger_{i,\alpha}c_{i,\alpha}$ counts the number of electrons in orbital $i$. As shown in Fig.~\ref{figS4}, $\varepsilon_i^*$ describes the gate-tunable chemical potential of each orbital. $U_i$ and $K_{ij}$ capture, respectively, intra- and inter-orbital Coulomb interaction energies. The term proportional to $\xi$ describes the spin correlation energy of the middle dot, favoring $S=1$ triplet configurations when both orbitals $1$ and $2$ are occupied. $\hat{S}$ is the total spin operator for the middle dot where spin in each orientation $\ell$ is given by $\hat{S}^\ell = \frac{1}{2}\sum_{\lambda,\alpha,\alpha^\prime}c^\dagger_{\lambda,\alpha}\sigma_{\alpha,\alpha^\prime}c_{\lambda,\alpha^\prime}$, summed over the orbitals $\lambda=1,2$. The final term in the Hamiltonian denotes tunnel-couplings $t_{ij}$ between orbitals $\langle i,j\rangle$ located in adjacent dots.

We implement specific multi-dot voltage pulses to explore the regimes of the effective exchange interaction, so we rewrite the orbital parameters (shown in Fig.~\ref{figS4}) as
\begin{multline}
\label{eqnS3}
\varepsilon_S = \varepsilon_2^* - \varepsilon_1^*, \quad
\bar{\varepsilon} = \frac{\varepsilon_L^*+\varepsilon_1^*+\varepsilon_R^*}{3}, \\
\varepsilon^* = \frac{\varepsilon_L^*-\varepsilon_R^*}{2} \quad
\varepsilon_M^* = \varepsilon_1^* - \frac{\varepsilon_L^*+\varepsilon_R^*}{2}.
\end{multline}
The first term, $\varepsilon_S$, is the spacing between the first and second orbitals of the middle dot. In our model, $\varepsilon_S$ is determined by the mesoscopic details of the dot and is independent of the plunger gate voltage $V_M$, so we take this as a fixed parameter. Tuning the second term, $\bar{\varepsilon}$, while keeping all others in Eqn.~\ref{eqnS3} constant is equivalent to a uniform voltage pulse on all dots, so we neglect it. The next term, $\varepsilon^*$, sets the difference between the chemical potentials of the left and right dots. This is proportional to the gate voltage $\varepsilon$, up to some lever arm factor, in addition to a factor of $1/\sqrt{2}$ arising from a difference in definition. The last term, $\varepsilon_M^*$, controls the detuning of the middle dot chemical potential relative to the left and right dots. This is proportional to the gate voltage $V_M$ again up to some lever arm factor. We have reduced the four gate-tunable chemical potentials $\varepsilon_i$ in our Hubbard model to two variables, $\varepsilon^*$ and $\varepsilon_M^*$, which will affect exchange. These terms are, respectively, the $x$ and $y$ axes of Fig.~\ref{fig3}c and \ref{figS5}.

There are numerous parameters in the Hubbard model whose values need to be fixed to calculate exchange oscillations for Fig.~\ref{fig3}c. However, exact values for most terms, particularly the various Coulomb energies, $U_i$ and $K_{ij}$, are not known. Fortunately, the charging energy of the multielectron dot is known to be approximately $1\:\mathrm{meV}$. Therefore, we employ some simplifying assumptions to reduce the number of parameters in the model, summarized in Table~\ref{tabS2}. We expect the middle-dot intra-orbital Coulomb interaction energies $U_1$ and $U_2$ and the inter-orbital Coulomb interaction energy $K_{12}$ to be comparable, so we assume that they are equal. This is convenient, as it allows us to define an energy scale $U$, approximately equal to the $1\:\mathrm{meV}$ charging energy of the middle dot and proceed by defining the remaining Coulomb terms relative to $U$. The left and right dots are smaller than the middle dot, and are expected to have an appropriately larger intra-orbital Coulomb interaction energy. Based on their size relative to the middle dot, we denote $U_L = U_R \approx 5U$. The remaining $K_{ij}$ terms have been estimated based on the spacing between dots in the device; nearest-neighbor terms are assumed equal and $0.10\:U$, while the next-nearest neighbor term $K_{LR}$ is $0.02\:U$. We have set the spin correlation energy $\xi$ based on experiments in Ref.~\onlinecite{Malinowski2018} conducted on the device on the same chip. Since the Coulomb energies of the left and right dots are large with respect to the middle dot, our model will only consider tunneling events \emph{onto} the middle dot. Furthermore, we shall assume that tunneling terms from the left and right dots to the same orbital are equal, such that  the model reduces to two tunneling parameters: $t_1$ and $t_2$. The last three terms in Table~\ref{tabS2} ($\varepsilon_S$, $t_1$ and $t_2$) represent tuning parameters for the exchange mediated by the middle dot whose values have been chosen to reproduce an oscillation pattern qualitatively similar those in Fig.~\ref{fig3}b.

\begin{table}
	\caption{Summary of the values of parameters for the Hubbard Hamiltonian for Fig.~\ref{fig3}c. The intra-dot Coulomb interaction energy $U$ has been used to define an energy scale to estimate the remaining parameters.}
	\label{tabS2}
	\begin{tabular}{c|c}
		 & \textbf{Value} \\
		\textbf{Parameter} & \textbf{(relative to $U$)} \\
		 \hline
		 $U_1 = U_2 = K_{12} \equiv U$ & $1.00$ \\
		 $U_L = U_R$ & $5.00$ \\ 
		 $K_{L1} = K_{L2} = K_{1R} = K_{2R}$  & $0.10$ \\
		 $K_{LR}$ & $0.02$ \\
		 $\xi$ & $0.10$ \\
		 $\varepsilon_S$ & $0.06$ \\
		 $t_{L1} = t_{1R} \equiv t_1$ & $0.04$ \\
		 $t_{L2} = t = {2R} \equiv t_2$ & $0.01$ \\
	\end{tabular} 
\end{table}
           
The Hamiltonian in Eqn.~\ref{eqnS2} can be solved in the two-electron regime to extract the gate-dependent effective exchange splitting $J_{\mathrm{eff}}(\varepsilon^*,\varepsilon_M^*)$ between the singlet and triplet states of the two spins. When the middle dot is far detuned from the left and right dots there is no exchange and the eigenstates of the Hamiltonian are $\ket{\uparrow\downarrow}$ and $\ket{\downarrow\uparrow}$. A mediated exchange interaction is induced by applying gate voltages affecting $\varepsilon^*$ and $\varepsilon_M^*$, resulting in flip-flops between the two electronic spins.
In the simulations, we track oscillations between $\ket{\uparrow\downarrow}$ and $\ket{\downarrow\uparrow}$ by detecting the spin states of both double quantum dots via spin-to-charge conversion. This is different from the experiment where the precession occurs between $\ket{S_L}\ket{S_R}$ and $\frac{1}{2}(\ket{S^L}\ket{S^R}-\ket{T_0^L}\ket{T_0^R}+\ket{T_+^L}\ket{T_-^R}+\ket{T_-^L}\ket{T_+^R})$ states. However, the observed pattern of oscillations is the same, except for the visibility (100 \% visibility in the simulations and expected 75\% visibility in the experiment), which is adjusted manually in the presented simulations. The oscillations in Fig.~\ref{fig3}c, calculated from $J_{\mathrm{eff}}(\varepsilon^*,\varepsilon_M^*)$, describe the probability of recovering an initial state $\ket{\uparrow\downarrow}$ after an evolution time $\tau = 6\:\mathrm{ns}$, for a set of gate voltages $(\varepsilon^*,\varepsilon_M^*)$. The range of the $\varepsilon^*$ and $\varepsilon_M^*$ axes have been chosen to cover electronic configurations (1,2$N$,1), (0,2$N$+1,1), (1,2$N$+1,0) and (0,2$N$+2,0), as in Fig.~\ref{fig3}c. Recall that the `unoccupied' state of the middle dot describes an effective vacuum with 2$N$ electrons. The evolution time $\tau$ for the Hamiltonian has been estimated using the approximation $U \approx 1\:\mathrm{meV}$. Simulations in Fig.~\ref{fig3}c qualitatively reproduce the three regimes of exchange interaction observed in Fig.~\ref{fig3}b.

In (0,2$N$+2,0), the middle dot has two excess electrons, one from each adjacent dot. There are three possible two-electron states for the middle dot. There may be two electrons singlet-paired in the lowest orbital, or one electron in each orbital, forming either a triplet or singlet state. Neglecting Coulomb interactions, the latter two states are gapped from the former by $\varepsilon_S$ when $\xi=0$. However, the presence of the non-zero spin correlation term in Eqn.~\ref{eqnS2} lowers the energy of the triplet state by  $\xi$. We have tuned $\varepsilon_S < \xi$ such that the two-electron ground state is a triplet spin configuration. Thus, in this region the two electrons, located in the same dot, have an `onsite' exchange splitting $J_{\mathrm{eff}}\approx\varepsilon_S-\xi$, which is negative (triplet-favoring) and small, producing rapid oscillations. 

For (0,2$N$+1,1)/(1,2$N$+1,0) the middle dot has an excess electron from the left/right dot. The electrons are now located in adjacent sites, so `direct' exchange interaction arises from virtual occupation of the middle dot. A perturbative analysis demonstrates the exchange splitting will have the generic form $J_\mathrm{eff}\approx {2t_1^2}/{\Delta_S}-{t_2^2}/{\Delta_T}$. 
The symbols $t_{1/2}$ are the tunneling terms from earlier, while $\Delta_{S/T}$ describe the energy costs of an electron tunneling into either the first orbital and forming a singlet state ($\Delta_S$), or tunneling into the second orbital and forming a triplet state ($\Delta_T$). These $\Delta$ terms are linear combinations of gate voltages from Eqn.~\ref{eqnS3}, Coulomb interaction energies and the spin correlation energy. Note that $J_\mathrm{eff}$ may be positive (singlet-favoring) or negative (triplet-favoring) depending on the choice of parameters. Following the previous section, since $U_1 = K_{12}$ and $\varepsilon_S<\xi$, we expect $\Delta_T < \Delta_S$. However, when the middle dot is far detuned from the left and right dots, $\Delta_T \approx \Delta_S$ and the overall sign of $J_{\mathrm{eff}}$ is determined by the ratio of the tunneling terms. We set $t_1 > t_2/ \sqrt{2}$ such that, in this detuned regime, $J_\mathrm{eff}$ is positive (singlet-favoring), opposite to the previous region. This splitting grows more positive as we tune towards (0,2$N$+2,0), producing more rapid oscillations. Critically, near the charge transition between (1,2$N$+1,0)/(0,2$N$+1,1) and (0,2$N$+2,0), we see a maximum in the exchange profile $J_{\mathrm{eff}}(\varepsilon^*,\varepsilon_M^*)$ as the dominant source of exchange changes from singlet-favoring to triplet-favoring. This maximum produces chevrons in the oscillation pattern. In Fig.~\ref{fig3}c charge noise on the gates results in blurring that is larger than the fringe separation, obscuring the chevrons. However, they are evident in Fig.~\ref{fig4}d, taken at a shorter evolution time.

Finally, in (1,2$N$,1), the electrons occupy distant dots and the exchange splitting approaches zero. However, there is still a weak interaction which arises from virtual occupation of the middle dot. Again, the sign of this exchange is positive since we have set $t_1 > t_2/ \sqrt{2}$. However, the process involves co-tunneling events from the left and right dots. Therefore $J_{\mathrm{eff}}\propto t_1^4$ and the splitting is much smaller, eventually vanishing as the middle-dot orbitals are detuned even further. The oscillations in this region are correspondingly slower and eventually vanish. 

\section{Effects arising from finite rise time of apparatus}
\label{risetime}

%:fig5
\begin{figure}
	\centering
	\includegraphics[width=0.48\textwidth]{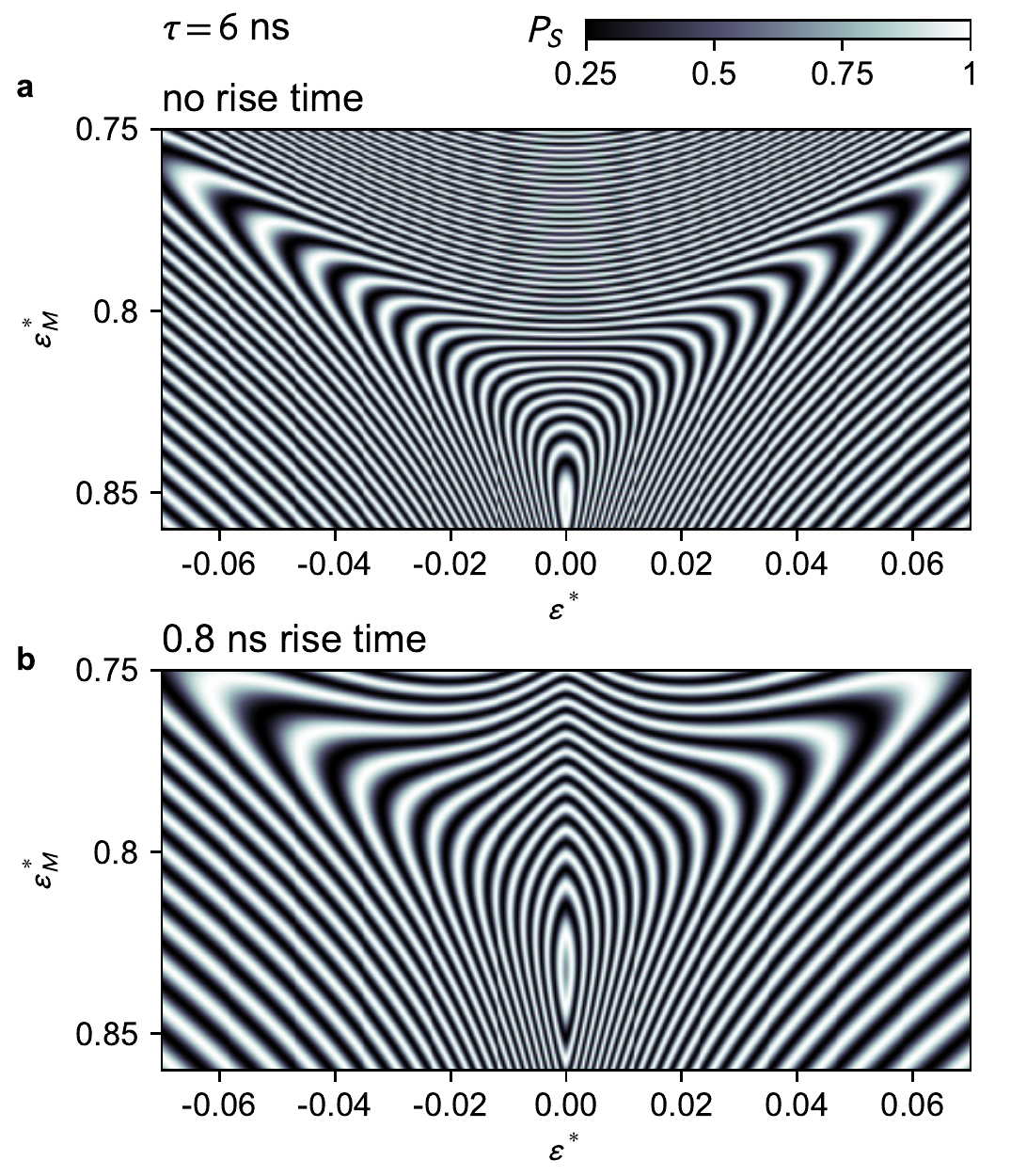}
	\caption{
	Comparison of the simulated fingerprint at the crossover between direct and onsite exchange regimes, excluding (a) and  including (b) effects of a finite rise time.
	}
	\label{figS5}
\end{figure}

Due to the high frequency of the exchange oscillations and the short duration of the exchange pulses, the rise time of our apparatus has pronounced distorting effects on the observed oscillation patterns. A simulation based on square pulses would not be realistic, and hence we assume a simple phenomenological model for the time dependence of $\varepsilon_M^*$ used in our simulations:
\begin{equation}
	\varepsilon_M^* (t) = \tilde{\varepsilon}_M^* - (1.5 + \tilde{\varepsilon}_M^* e^{-t/\tau_0})
	\label{eqnS5}
\end{equation}
where $\tau_0=0.8$~ns, $\tilde{\varepsilon}_M^*$ is the value displayed on the vertical axis in the figures presenting the simulations and $\varepsilon_M^* (0) = -1.5$ is chosen for convenience, but its precise value has no qualitative influence on the obtained oscillations pattern.

We expect that  finite-rise-time effects are most pronounced in the $\varepsilon$-$\MP$ fringe pattern obtained with short interaction time $\tau=2$~ns, presented in Fig.~\ref{fig4}a.
Our simple model in Eq.~\ref{eqnS5} captures most, but not all distortion effects observed in this data. For example,  Fig.~\ref{figS5} shows that a finite rise time can result in the upward bending of fringes around the symmetry point in the onsite exchange interaction regime, similar to what is observed in Fig.~\ref{fig4}a. This effect can also be understood intuitively, by noting that the smaller $|\varepsilon|$ is, the larger is the exchange interaction $J$ while $V_M$ is rising. However, some differences between simulations and experiment remain, and may require gate-voltage-dependent parameters in the Hamiltonian of Eq.~\ref{eqnS3} and a more realistic waveform in Eq.~\ref{eqnS5}.

\end{document}